\documentclass{article}
\usepackage{spconf,amsmath,graphicx,hyperref, booktabs, multirow, color}
\usepackage{amssymb}
\usepackage{arydshln}
\usepackage{enumitem}

\title{Controllable Embedding Transformation \\ for Mood-Guided Music Retrieval}
\name{Julia Wilkins$^{1,2}$,
      Jaehun Kim$^{1}$,
      Matthew E. P. Davies$^{1}$,
      Juan Pablo Bello$^{2}$,
      Matthew C. McCallum$^{1}$
      }
\address{
$^{1}$SiriusXM-Pandora, USA \;
$^{2}$New York University, New York, USA 
}
\begin{document}
\ninept
\maketitle
\begin{abstract}
Music representations are the backbone of modern recommendation systems, powering playlist generation, similarity search, and personalized discovery. Yet most embeddings offer little control for adjusting a single musical attribute, e.g., changing only the mood of a track while preserving its genre or instrumentation. 
In this work, we address the problem of controllable music retrieval through embedding-based transformation, where the objective is to retrieve songs that remain similar to a seed track but are modified along one chosen dimension. 
We propose a novel framework for mood-guided music embedding transformation, which learns a mapping from a seed audio embedding to a target embedding guided by mood labels, while preserving other musical attributes. Because mood cannot be directly altered in the seed audio, we introduce a sampling mechanism that retrieves proxy targets to balance diversity with similarity to the seed. 
We train a lightweight translation model using this sampling strategy and introduce a novel joint objective that encourages transformation and information preservation. Extensive experiments on two datasets show strong mood transformation performance while retaining genre and instrumentation far better than training-free baselines, establishing controllable embedding transformation as a promising paradigm for personalized music retrieval. 
\end{abstract}
\section{Introduction}
\label{sec:intro}

Music consumption behavior on streaming platforms can range from passive background listening to active playlist creation and explicit recommendation feedback \cite{schedl2021music}. A promising direction within this continuum targets the discovery of music which shares many underlying musical properties of some seed track(s), but differs in one or two targeted dimensions. While most modern music recommendation systems are built upon learned music representations that capture high-level semantic properties and enable efficient retrieval across millions of tracks  \cite{schedl2021music, schedl2018current, reccTrends, deldjoo2024content}, these representations lack mechanisms for fine-grained control of specific properties such as mood or genre. In this study, we focus specifically on musical mood, addressing a novel use case to enable listeners to guide the retrieval of content which is e.g., ``\textit{similar, but happier},'' or ``\textit{similar, but more energetic}''.

Music style transfer has been well-explored in generative frameworks, where models learn to alter a song given a guidance signal (e.g., emotion or instrumentation) while preserving core musical content \cite{dai2018music, li2024music, hu2020make, copet2023simple}. These approaches are useful for small-scale creative applications \cite{nistal2024diffariffmusicalaccompanimentcocreation, cifka2020groove2groove} but less so in a retrieval setting at scale due to costly computation associated with generating new audio. Recent works have begun to bridge the gap between  generative and representation learning approaches for style transfer and retrieval tasks; \cite{chu2025text2fx} leverages an audio-text embedding space to manipulate audio effects using natural language prompts, and \cite{guinot2025gdretrievercontrollablegenerativetextmusic} uses diffusion to generate audio queries conditioned on text for text-music retrieval.

There is little prior work on manipulating music embeddings in an audio-only latent space for semantically guided retrieval tasks. Disentanglement-based approaches learn separate latent subspaces for distinct musical attributes to use in targeted downstream applications \cite{guinot2024leaveoneequivariantalleviatinginvariancerelatedinformation, lee2020disentangledmultidimensionalmetriclearning, wilkins2025balancing, tanaka2025unsupervised}, but do not allow for manipulation of an input audio embedding along a specific axis within these subspaces. The most closely related work is \cite{mccallum2024similarfastermanipulationtempo}, where the model learns a tempo-controlled latent transformation that can be used to retrieve tracks that are similar but of a different tempo. In this approach, translated target embeddings are generated by directly modifying the tempo of the input audio signal. While this demonstrates the potential of embedding space manipulations for guided musical retrieval, for more abstract musical attributes such as mood, directly transforming the input signal to generate positive training pairs is non-trivial.


To this end, in this work we introduce the task of mood-guided embedding transformation for music retrieval. Our contributions are:
\begin{itemize}[leftmargin=*, itemsep=0.3em]
\item We present a controllable music embedding transformation framework that translates a seed track to an embedding aligned with a target mood, enabling track retrieval in the new mood while preserving other musical qualities (e.g., genre and instrumentation).
\item We introduce a nearest-neighbor sampling scheme that yields seed–target pairs differing in mood but otherwise similar, for cases where signal-based augmentations are infeasible.
\item We design complementary loss functions for training a lightweight transformation model and provide an extensive ablation study highlighting their individual contributions.
\item We show that our approach significantly outperforms baselines in mood transformation and information preservation on both a large-scale proprietary dataset and MTG-Jamendo \cite{bogdanov2019mtg}.

\end{itemize}
\begin{figure*}
    \centering 
    \includegraphics[width=0.95\linewidth, trim={1.3cm 1.5cm 1.3cm 2cm}]{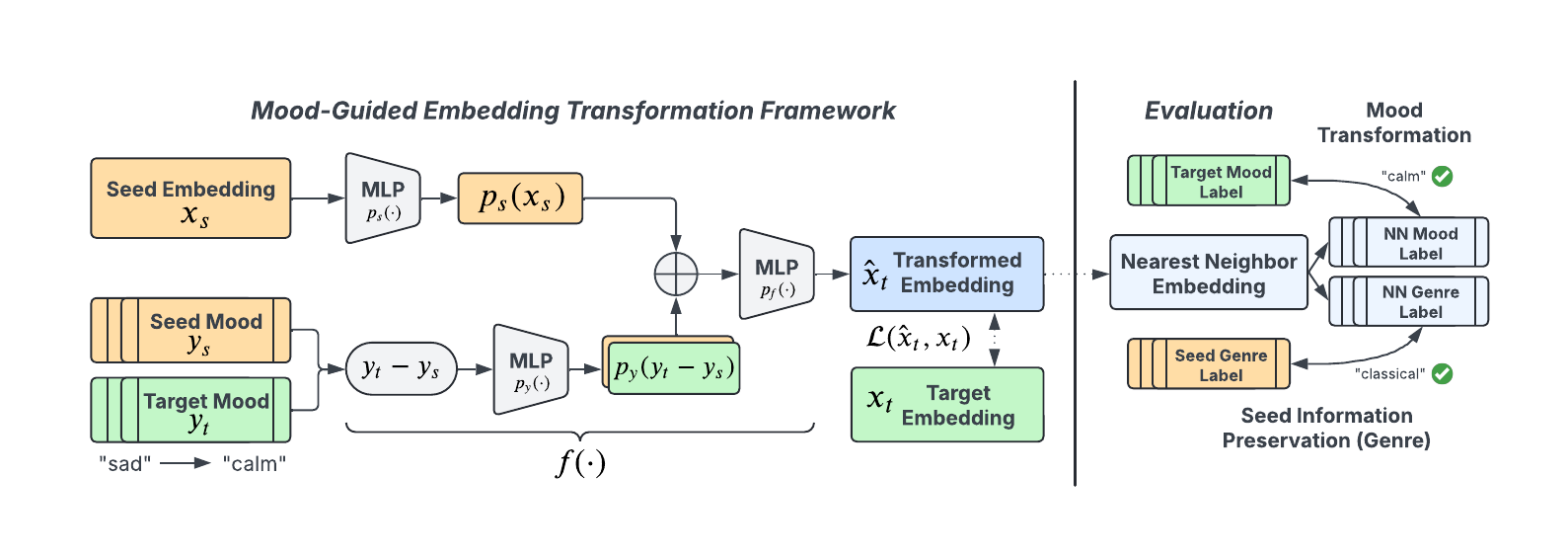}
    \caption{
    A MULE embedding of a seed song is transformed, guided by a target mood label, into an embedding that can retrieve similar songs modified along the mood dimension. We evaluate in terms of mood transformation and seed information preservation. 
    }
    \label{fig:main}
\end{figure*}

\vspace{-0.3cm}
\section{Method}
\label{sec:method}
We propose a novel framework for controllable music embedding transformation. The goal of our system is to learn a transformation purely in the embedding space that shifts a single, controllable attribute of an input audio track, while preserving other musical attributes. We use \textit{mood} as the transformation attribute and \textit{genre} and \textit{instrumentation} as measurable attributes that should be preserved. 

We use ``seed'' throughout to refer to the original song and its associated embedding and musical attributes, and ``target'' as the desired mood and an associated song embedding 
to use as the transformation goal. We operate within the open-source MULE \cite{mccallum2022supervisedunsupervisedlearningaudio} embedding space for this study,
which has shown state-of-the-art (self-supervised) performance on musical representation learning tasks including mood and genre prediction, and thus provides a suitable foundation for embedding-based transformation.


\vspace{-0.3cm}
\subsection{Mood-Guided Embedding Transformation Framework}
Our method learns a mapping $f(\cdot)$ from a MULE embedding of a seed audio track to the embedding of a target audio track of a different mood, while preserving attributes \textit{besides} mood in the seed track. To this end, we must select target embeddings that represent a mood-adjusted version of the seed. Because we cannot directly manipulate the mood of the input track to generate the true mood-adjusted embedding, we introduce a novel data sampling mechanism to retrieve proxy target embeddings to use in training:

Let $\mathbf{x_s} \in \mathbb{R}^d$ denote the MULE embedding of a seed audio track with mood label $\mathbf{y_s} \in \mathcal{M}^{m}$, where $d$ is the embedding dimension, $\mathcal{M}$ is the set of possible moods, and $m$ is the dimension of the one-hot mood label vectors. For every seed, we compute the top-$100$ most similar tracks per mood to that track via cosine similarity in the MULE embedding space. We store this mapping and use it to draw target tracks at training time. For each seed $\mathbf{x_s}$, we first select a target mood $\mathbf{y_t} \in \mathcal{M}^m$ at random. Given $\mathbf{y_t}$ and $\mathbf{x_s}$, we sample a target track embedding $\mathbf{x_t}$ from the similarity map. To maximize diversity of tracks included and avoid embedding hubs, we sample target tracks at random from the top-100 most-similar tracks per seed. When the seed and target moods match, the seed embedding is used as the target to incentivize the model to learn an identity mapping.

The full training framework is shown in Figure \ref{fig:main}. We use the sampling mechanism to gather three inputs to the model for training: (1) seed track embedding ($\mathbf{x_s}$), (2) seed mood ($\mathbf{y_s}$), encoded as a one-hot label vector, and (3) target mood ($\mathbf{y_t}$), similarly encoded. The MULE embeddings ($\mathbf{x_s}$, $\mathbf{x_t}$, $\mathbf{\hat{x}_t}$) have dimension $d$\,=\,$1728$, and the one-hot encoded mood label vectors, $m$\,=\,$4$. The seed embedding is projected through an MLP ($p_s(\cdot)$) to a $512$-dimensional vector. We subtract the seed mood vector from the target mood vector as a guidance signal and project this through a separate MLP $p_y(\cdot)$ to a $128$-dimensional vector to increase the capacity of the guidance signal so that the dimensionality of the seed embedding does not dominate. The projected seed and label difference vectors are concatenated before passing through $p_f(\cdot)$ to move back to the target dimension of $1728$. The model output is the transformed embedding, $\mathbf{\hat{x}_t} = f(\mathbf{x_s}, \mathbf{y_s}, \mathbf{y_t})$, which is then compared to the target mood-adjusted embedding ($\mathbf{x_t}$) through a joint objective described next.

\subsection{Objective Design}\label{sec:obj_design}
To ensure that the transformed embedding is aligned with the target mood while preserving properties of the seed, we use three complementary loss terms. \textbf{Cosine similarity}: we encourage the predicted embedding $\mathbf{\hat{x}_t}$ to be similar to the target embedding $\mathbf{x_t}$ by maximizing their cosine similarity, thus minimizing $1$ minus this quantity:

\vspace{-0.2cm}
\begin{equation*}
    \mathcal{L}_{\text{cosine}} = \frac{1}{B}\sum_{i=1}^{B}(1-cos(\mathbf{\hat{x}_t}^{(i)}, \mathbf{x_t}^{(i)})),
\end{equation*}
where $B$ is the batch size and $\mathbf{\hat{x}_t}$ and $\mathbf{x_t}$ are normalized.

\textbf{Triplet loss}: to ensure that the transformed embedding is close to the target, but distinct from the seed embedding, we use a triplet-style hinge loss \cite{schroff2015facenet, lee2020disentangledmultidimensionalmetriclearning}. For each sample, the transformed embedding $\mathbf{\hat{x}_t}$ is considered the ``anchor'', the target embedding $\mathbf{x_t}$ is the positive, and the seed embedding $\mathbf{x_s}$, is the negative:

\vspace{-0.2cm}

\begin{equation*}
\mathcal{L}_{\text{triplet}} = \frac{1}{B} \sum_{i=1}^B
\max \Bigg( 0, \; \alpha + \cos(\hat{\mathbf{x}}_t^{(i)}, \mathbf{x}_s^{(i)}) - \cos(\hat{\mathbf{x}}_t^{(i)}, \mathbf{x}_t^{(i)}) \Bigg),
\end{equation*}
where $\alpha$ is a margin hyperparameter. This loss penalizes similarity between the transformed embedding and the seed, while rewarding similarity between the transformed and target embeddings, explicitly encouraging movement in the embedding space.

\textbf{Cosine BCE}: we use a contrastive-style loss \cite{chen2020simple} that emphasizes the difference between transformations when the seed and target moods are different vs. the same. This is a more nuanced version of the vanilla $\mathcal{L}_{\text{cosine}}$ above which is label-agnostic. 
This loss acts as a label-dependent regularizer that emphasizes learning the identity mapping when $\mathbf{y_s} = \mathbf{y_t}$, and a relaxation of this alignment when $\mathbf{y_s} \neq \mathbf{y_t}$ via binary cross-entropy targets of $1$ and $0.5$, respectively. 


\vspace{-0.2cm}

\begin{equation*}
    \mathcal{L}_{\text{cosBCE}} = \frac{1}{B} \sum_{i=1}^{B}
\text{BCE}\Big( \sigma \big( \gamma \cdot \cos(\hat{\mathbf{x}}_t^{(i)}, \mathbf{x}_t^{(i)}) \big), \; t^{(i)} \Big),
\end{equation*}
where $\sigma$ is the sigmoid function, $\gamma$ is a scaling factor, and $t^{(i)} \in \{1, 0.5 \}$ is the BCE target that depends on mood match.



The full training objective is the weighted sum of the above: $\mathcal{L} = 
\lambda_{\text{cosine}} \, \mathcal{L}_{\text{cosine}} \;+\;
\lambda_{\text{triplet}} \, \mathcal{L}_{\text{triplet}} \;+\;
\lambda_{\text{cosBCE}} \, \mathcal{L}_{\text{cosBCE}}$.


\section{Experimental Design}
\label{sec:expdesign}

\subsection{Datasets}
We use a large-scale proprietary music dataset for our study that contains $1.3$\,M songs with high-quality mood and genre annotations. This dataset contains songs from a set of four moods pertaining to high and low-energy and positive and negative sentiment, which approximately align with the main dimensions of Russell's valence-arousal model \cite{russell1980circumplex}, differing from the more widely used quadrants. The dataset also has genre annotations spanning $20$ classes. As a secondary, publicly-available dataset, we use MTG-Jamendo \cite{bogdanov2019mtg}. In the absence of an exact match in mood taxonomies, we use the subset of the best matching mood labels, namely: ``energetic'', ``calm'', ``happy'',  and  ``sad'', totaling $4$\,k full-length songs. MTG-Jamendo provides good-quality audio recordings, but the annotations are derived from user tags which can inherently be noisy. In addition to mood and $94$-class genre labels, MTG-Jamendo has multi-label instrumentation labels across $40$ instrument categories, allowing us to examine an additional property of the transformed embeddings, independent of mood and genre.

For both datasets, if tracks are multi-labeled in mood or genre, we choose a single label at random per track.  For training splits, we enforce disjointness at the artist level to prevent leakage and stratify by mood across splits. We use an $80/10/10$ training/validation/test split. We compute the MULE \cite{mccallum2022supervisedunsupervisedlearningaudio} embeddings as per the open-source implementation, taking $3$\,s windows, each comprised of $300$ $96$-band Mel-Spectrogram frames centered $2$\,s apart, resulting in overlapping context windows. The resulting embedding timeline is then averaged to form a single embedding per song. For more details on the audio analysis parameters, see \cite{mccallum2022supervisedunsupervisedlearningaudio} and its accompanying codebase.

\subsection{Training Setup}
The projection modules used in training are 2-layer MLPs with ReLU activations. The seed projector $p_s(\cdot)$ has a hidden layer of dimension $1024$ and an output layer of dimension $512$, and the label projector $p_y(\cdot)$ has a hidden dimension of $64$ and an output dimension of $128$. For both $p_s$ and $p_y$, we apply dropout after the hidden layer with rates of $0.3$ and $0.4$, respectively. The final projector $p_f(\cdot)$, applied to the concatenation of $p_s(x_s)$ and $p_y(y_t - y_s)$, consists of a single linear layer with dropout $0.3$ before the output projection back to the original MULE dimensionality of $1728$.

To understand the contribution of each loss component and select the optimal training configuration, we conduct an ablation study across both datasets for our three objectives described in Sec.~\ref{sec:obj_design}. We perform full training and evaluation runs on both datasets for each loss term individually and in combination with each other term with the $\lambda$ weighting set to either $1$ or $0$ if the loss is included or not. We choose the best loss configuration based on a weighted average of the validation mood and genre evaluation metrics described in Sec.~\ref{ssec:metrics} and use this as our best model per dataset, where we place a slightly higher weight on the mood component. In our final training configuration all $\lambda$ weights are set to $1$, $\alpha = 0.3$ in $\mathcal{L}_{\text{triplet}}$, and $\gamma=3$ in $\mathcal{L}_{\text{cosBCE}}$.

Models trained on the large dataset are trained for $100$ epochs using a batch size of $1024$ and AdamW optimization with a linear learning rate of $1e-5$. MTG-Jamendo models are trained for $500$ epochs with a batch size of $1024$, AdamW, and a linear learning rate of $5e-4$. Learning rates are determined through hyperparameter tuning during the loss ablation. For the MTG-Jamendo models, due to the small scale of the dataset, we use $3$-fold cross validation (random $8$:$1$:$1$ splits within each fold) and report metrics as an average of folds across test sets for all results including baselines. We perform model selection based on the best mood precision score in validation.


\subsection{Evaluation Metrics}
\label{ssec:metrics}
The two core qualities we measure are (1) mood transformation, and (2) seed information preservation, evaluated via genre and instrumentation in this study. In the absence of associated labels for the transformed embedding, we examine the labels of the nearest neighbor embeddings. For mood transformation, we evaluate the mood label of the nearest neighbor to the transformed embedding versus the target mood label and measure precision at 1, denoted \textbf{Mood P@1}. For genre information preservation, we compare the nearest neighbor genre label of the transformed embedding to the seed genre label to measure seed-genre precision at 1, denoted \textbf{Genre P@1}. For the multi-label instrumentation consistency evaluation, we use Jaccard score 
between the nearest neighbor instrumentation labels of the transformed embedding versus the seed labels, and denote this as \textbf{Inst. J@1}.


\begin{figure*}[t]
\vspace{-0.3cm}
  \centering
  \begin{minipage}[t]{0.49\linewidth}
    \centering
    \includegraphics[width=\linewidth,height=3.8cm,trim={0cm 3cm 0 3cm}]{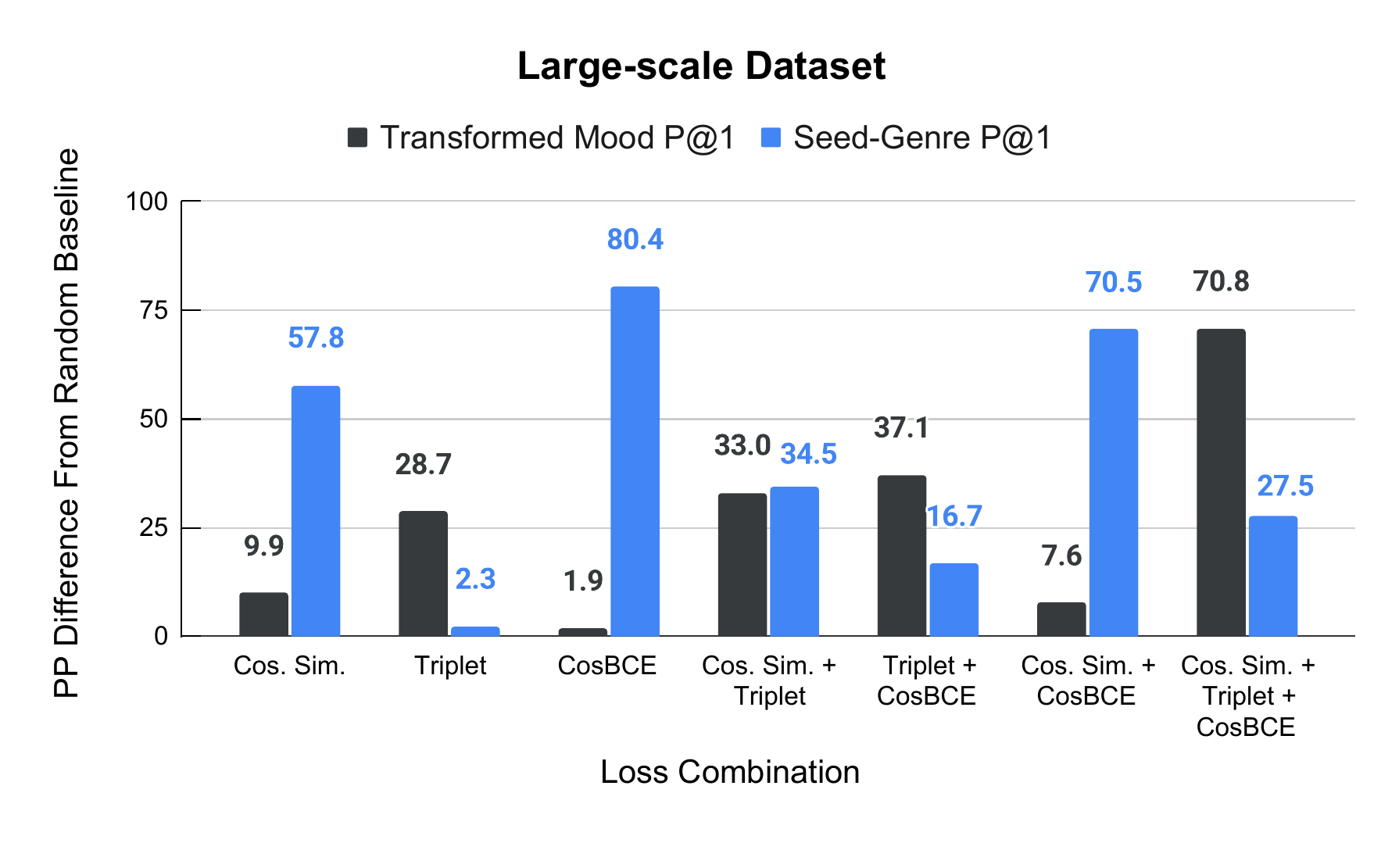}
    \label{fig:jamendo_bar}
  \end{minipage}\hfill
  \begin{minipage}[t]{0.49\linewidth}
    \centering
    \includegraphics[width=\linewidth,height=4cm,
    trim={0 3cm 0 2cm}]{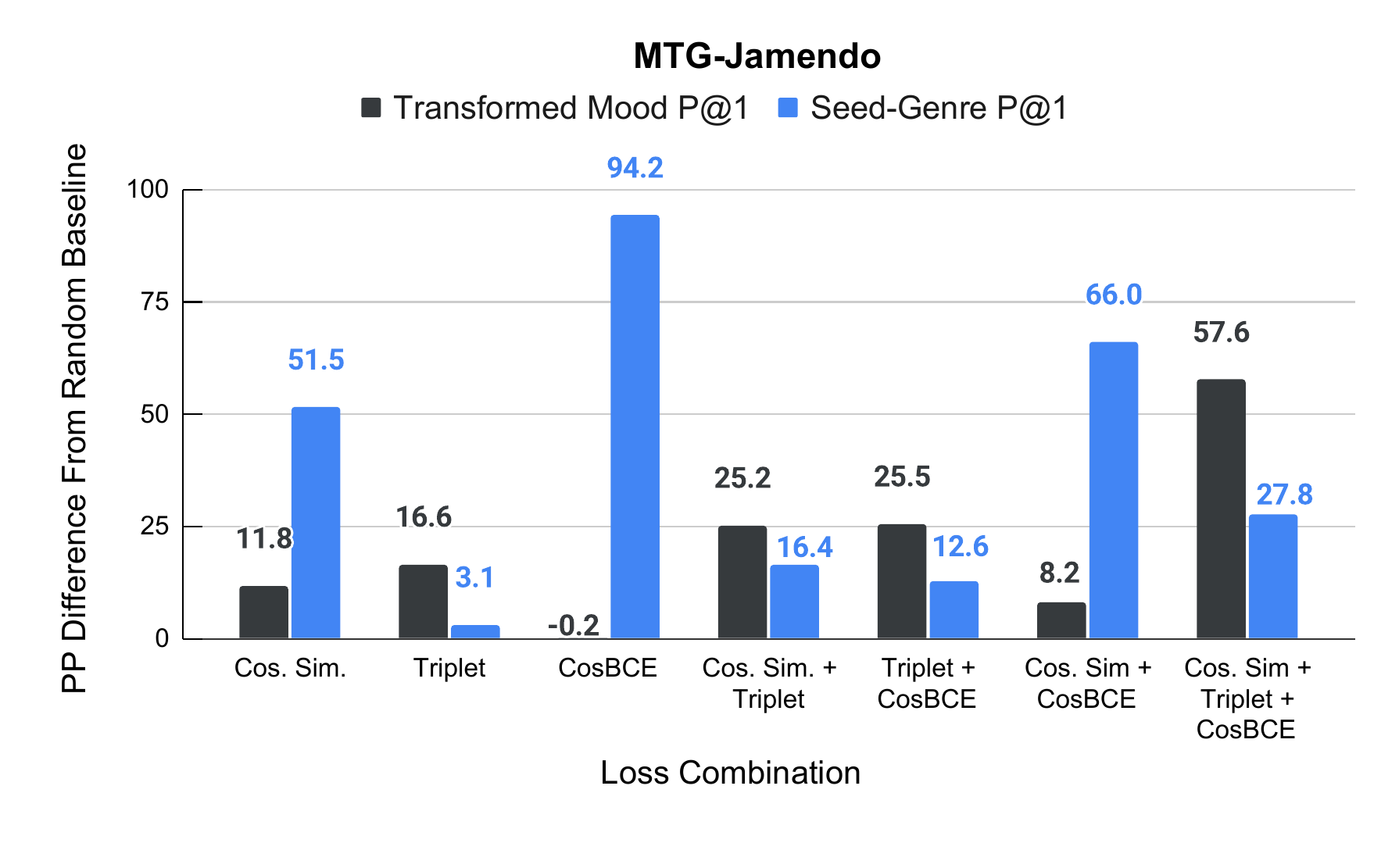}
  \end{minipage}
  \caption{Loss ablation using the large-scale dataset and MTG-Jamendo. Results shown are in percentage-points (pp) difference from the random baseline, scaled by $100$. For both datasets, the random baseline for Mood P@$1$ is $0.25$. For Genre P@$1$, the baseline is $0.05$ for the large-scale dataset and $0.011$ for MTG-Jamendo.}
  \label{fig:combined_bar}
\end{figure*}

\begin{table}[h]
\vspace{-0.2cm}
\centering
\scriptsize
\caption{Main results on the large-scale dataset and MTG-Jamendo against baselines. We measure Mood P@1 to evaluate mood transformation and Genre P@1 and Inst. J@1 to ensure information from the seed embedding is preserved through the mood transformation.}
\vspace{0.2cm}
\setlength{\tabcolsep}{5pt} 
\begin{tabular}{@{}lccc@{}} 
\toprule
\multirow{2}{*}{\textbf{Large-scale Dataset}} 
& \textbf{Mood P@1} & \textbf{Genre P@1} &  \\
& $(N=4)$ & $(N=20)$ &  \\
\midrule
\textbf{Ours: $\mathcal{L}_{\text{cosine}} + \mathcal{L}_{\text{triplet}} + \mathcal{L}_{\text{cosBCE}}$  } & 0.96 & 0.32 &  \\
Baseline: Random & 0.25 & 0.05 \\
Baseline: Average mood vector  & 1.0 & 0.10 &  \\
Baseline: Oracle (Top-$1$) & 1.0 & 0.54 &  \\
Baseline: Oracle (Rand-$100$) & 1.0 & 0.38 &  \\
\midrule
\multirow{2}{*}{\textbf{MTG-Jamendo \cite{bogdanov2019mtg}}} 
& \textbf{Mood P@1} & \textbf{Genre P@1} & \textbf{Inst. J@1} \\
& $(N=4)$ & $(N=94)$ & $(N=40)$ \\
\midrule
\textbf{Ours: $\mathcal{L}_{\text{cosine}} + \mathcal{L}_{\text{triplet}} + \mathcal{L}_{\text{cosBCE}}$  }  & 0.83 &0.29 & 0.45 \\
Zero-shot from large-scale model  & 0.66 & 0.30 & 0.45\\
Fine-tuned on large-scale model & 0.68 & 0.29 & 0.41\\

Baseline: Random & 0.25 & 0.01 & 0.04\\
Baseline: Average mood vector & 0.82 & 0.05 & 0.18 \\
Baseline: Oracle (Top-$1$)  & 1.0 & 0.16  &  0.48 \\
Baseline: Oracle (Rand-$100$)  & 1.0 & 0.07  &  0.24 \\
\bottomrule
\end{tabular}
\label{tab:overall}
\end{table}

\vspace{-0.6cm}
\subsection{Baseline Methods}
We compare our system against the following training-free methods:

\textbf{Random}: as a lower bound, we provide a random baseline for Mood and Genre P@$1$ based on class cardinality. The Mood P@$1$ baseline is the same for both datasets as both contain $4$ moods, but MTG-Jamendo has a larger genre taxonomy than the large-scale dataset so the genre baseline is lower. We compute the random baseline for multi-label instrumentation Jaccard score in expectation, using an average of $2.77$ labels per sample in a Bernoulli formulation.

\textbf{Average mood embedding}: we compute the average vector (e.g., centroid) per mood using all embeddings in the test set. Given a seed embedding and a target mood, we use the target mood average vector as our ``generated'' embedding for evaluation. This version should, by design, be near-perfect for ``transformed'' Mood P@$1$, but lacks any notion of persisting seed information so genre and instrumentation scores should be low.




\textbf{Oracle}: this baseline uses our sampling mechanism directly to select a target embedding, instead of learning a transformed embedding to approximate the target as in our method. The Oracle represents an upper bound that requires access to dataset-wide mood labels or an accurate classifier to restrict candidates to the desired mood for inference. We report two Oracle settings: (i) \textit{Oracle Top-1}, which selects the most similar track to the seed within the target mood in an optimistic setting, and (ii) \textit{Oracle Rand-100}, which samples uniformly from the top-100 nearest target-mood candidates (matching our sampling exactly). Because Oracle candidates are chosen from the target mood by construction, Mood P@$1$ is trivially $1$. Genre and instrumentation scores should also exceed random and average baselines (especially for Top-1), since the chosen target is known to be close to the seed through the sampling mechanism.






\section{Results and Discussion}

\subsection{Core Results}

Our key results, shown in Table ~\ref{tab:overall}, demonstrate that our method consistently outperforms random baselines by a wide margin, achieving high mood transformation accuracy while simultaneously preserving genre and instrumentation.

On the large-scale dataset, our approach reaches Mood P@$1$ of $0.96$ and Genre P@$1$ of $0.32$, far exceeding random in both mood transformation and seed preservation. Against the average mood vector baseline, we show over a $3\times$ improvement in Genre P@$1$; this proves that our model is learning more nuanced transformations versus converging to the mood centroids alone. While the overly optimistic Oracle (Top-1) outperforms our system, our best method closely approximates the Oracle (Rand-100) in both metrics, illustrating that for a dataset with high-quality labeling, the nearest-neighbor sampling mechanism for training data pairs effectively provides an upper bound with regard to attribute (e.g., genre) preservation.
Further, the Oracle requires all labels at inference and the computation of mood similarity scores across all embeddings for a given target mood, whereas our model operates without labeled targets or similarity information at inference.


For MTG-Jamendo, in addition to improving upon the random baseline significantly across all metrics, our best model performs on par with the average mood baseline in terms of Mood P@$1$ while outperforming Genre P@$1$ and Inst. J@$1$ by $24$pp and $27$pp, respectively. Our strong results for instrumentation validate that our system preserves nuanced qualities of the input that are independent of the transformation axis. Surprisingly, our method improves upon both Oracle baselines in Genre P@$1$; we hypothesize that this may be due to the noisier labeling and the fact that the small yet diverse nature of the dataset leads to sparser embeddings in the MULE space, and thus lower similarity between embeddings within a certain mood. This would indicate that, despite an inherently weaker sampling mechanism for training with this dataset, our model is still able to learn a robust transformation.

We also show results for the large-scale dataset evaluated in a zero-shot manner on MTG-Jamendo, as well as a version that is fine-tuned on MTG-Jamendo. The zero-shot result is strong ($0.66$ vs. our best model at $0.83$ for Mood P@$1$), which may suggest both that our large-scale model can generalize relatively well to unseen data, and that MULE provides a robust backbone regardless of data shift. The lack of a significant improvement for fine-tuning versus zero-shot implies a domain gap between the datasets in terms of audio and labeling, but may be due to the small size of MTG-Jamendo.

\subsection{Loss Ablation Study} \label{sec:loss_ablation}

Figure \ref{fig:combined_bar} shows the outcome of our loss ablation, with results presented as percentage-point (pp) differences relative to the random baseline for each task. 
We observe generally consistent trends across both datasets, finding that relying on any single loss leads to relatively imbalanced outcomes. Using $\mathcal{L}_{\text{cosine}}$ alone provides moderate gains in genre preservation, but is limited in mood transformation. In contrast, the $\mathcal{L}_{\text{triplet}}$ drives improvements in mood transformation, as the only term explicitly incentivizing movement in the embedding space. Yet, this clearly comes at the expense of seed preservation in which metrics are close to random. $\mathcal{L}_{\text{cosBCE}}$ performs in the opposite way, achieving the highest Genre P@$1$ scores ($80.4$pp and $94.2$pp improvement from random on the large-scale and MTG-Jamendo datasets, respectively) but with negligible mood transformation.


Pairwise combinations of the losses partially mitigate these trade-offs, but the full combination of all three loss terms improves mood transformation the most, while still backed by a large increase in genre over random. On the large-scale dataset, the combined loss yields $70.8$pp and $27.5$pp improvements in Mood P@$1$ and Genre P@$1$, respectively, over the random baseline, while on MTG-Jamendo it achieves $55.7$ and $26.7$pp improvements in the same metrics. These results confirm that the objectives are complementary: cosine similarity encourages alignment, triplet loss enforces separation across moods, and cosine BCE acts as a stabilizer between the previous two terms. When used together they provide robust mood transformation while preserving seed information.
\section{Conclusion}
\label{sec:conclusion}
In this work, we introduce a framework for controllable music embedding transformation, enabling retrieval of tracks of a different mood but similar in other musical dimensions such as genre and instrumentation. We utilize a novel nearest-neighbor data sampling scheme to create seed-target embedding pairs to train our transformation model, and demonstrate in our evaluation on both a large-scale proprietary dataset and MTG-Jamendo that we are able to significantly outperform training-free baselines using this strategy. We show through an extensive loss ablation that success in both mood transformation and information preservation hinges on the design of complementary objective functions. In the future, we plan to investigate the interaction between multiple mood dimensions simultaneously (i.e., where a track can be both energetic and happy), and explore using an audio-text embedding space in our framework, where text embeddings could act as guidance instead of encoded labels.

\vfill\pagebreak


\bibliographystyle{IEEEbib}
\bibliography{refs}

\end{document}